\documentclass[epj]{svjour}
\usepackage{graphicx}
\usepackage{dcolumn}
\usepackage{bm}
\usepackage{amsmath,amssymb}

\newcommand{\n}{\normalfont~}

\newcommand{\bu}{B(E2)\negthinspace\protect\raisebox{0.035cm}{$\uparrow$} }
\newcommand{\buc}{B(E2)\negthinspace\protect\raisebox{0.035cm}{$\uparrow$}}
\newcommand{\parbu}{(B(E2)\negthinspace\protect\raisebox{0.035cm}{$\uparrow$}) }

\newcommand{\bua}{B(E2)\negthinspace\protect\raisebox{0.035cm}{$\uparrow$}$_{approx.}$ }

\begin{document}


\title{Seniority scenario for the $^{68-72}$Zn and $^{66-68}$Ni \bu difference}

\author{I. Deloncle \inst{1} \and B. Roussi\`ere \inst{2}}

\institute{CSNSM, IN2P3/CNRS and Universit\'e Paris-Sud, F-91405
Orsay Campus, France \and IPN, IN2P3/CNRS and Universit\'e
Paris-Sud, F-91406 Orsay Cedex,France}

\date{Received: date / Revised version: date}

\abstract{ In the seniority scheme the reduced
B(E2:\thickspace$0^+_1$$\rightarrow2^+_1$) \parbu transition probability of
even-even nuclei can be related to the product of the number of particles by
the number of holes of the valence space. This very simple expression is used
to analyze at the same time the experimental \bu values of $^{56-68}$Ni and
those of $^{62-72}$Zn. The evolution of these \bu values with neutron number
fits in with a scenario involving p-n interaction.}


\maketitle 
\section{Introduction\label{sec:intro}}
\vspace{-0.3cm}

The B(E2:\thickspace$0^+_1$$\rightarrow2^+_1$) reduced transition probability,
\buc, is correlated to the possibilities to get a $2^+_1$ state from
excitations in the single-particle spectrum underlying the $0^+_1$ one. Its
value is then very sensitive to the (sub-)shell structure. Low \bu values are
obtained for doubly closed-shell nuclei in which any excitation requires to
overcome the gap. High values are reached at mid-shell where the product of the
number of valence particles by the number of single-particle levels available
for the excitations is maximal. In order to study the shell structure of
neutron-rich nuclei with $N\simeq 40$, the \bu have been measured in the
$^{68}$Ni \cite{Sor02} and $^{72}$Zn nuclei \cite{Sor01} by coulomb excitation
experiments performed at Ganil. The results put light on an opposite behavior
of the \bu evolution above N=38 between the Ni and the Zn isotopes. This
difference is the subject of numerous research studies. Recently, experiments
concerning the \bu values in the two next neutron rich isotopes, $^{74}$Zn and
$^{70}$Ni have been performed at Ganil \cite{Sor02b} and at REX-ISOLDE
\cite{May02}, but their results are not yet available. Two theoretical papers
discussing the \bu of the Ni isotopes only in relation with a N=40 gap and its
size have been published \cite{Lan03,Cau04}. In the present work, on the basis
of simple calculations performed within the seniority framework and assuming
different (sub-)shell structure, we propose an overall interpretation of the
evolution of the \bu curves from the Ni up to the Zr isotopes. The p-n
interaction plays a large part in our scenario (see \cite{Del03}) which, in this
paper, is detailed for the Ni and Zn isotopes only. 

\vspace{-0.7cm}
\section{\label{sec:senior}A simple formula for \bu} 
\vspace{-0.3cm}

\begin{figure}[h] 
\hspace{0.25cm} 
\includegraphics*[height=5.5cm]{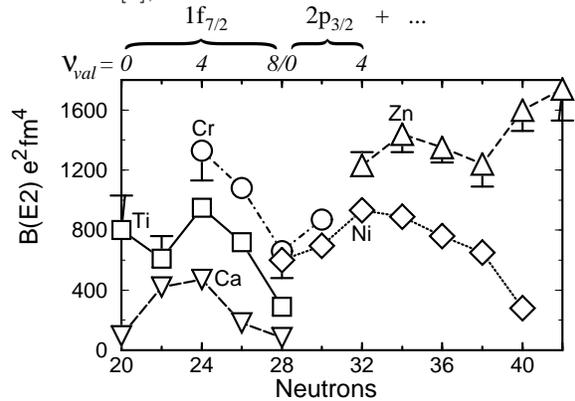} 
\vspace{-0.16cm}

\caption{Experimental \bu values of Ca, Ti, Cr, Ni and Zn isotopes from
\cite{Ram2}. Only one half of the error bar, when larger than the symbols, is
drawn. The number of valence neutrons ($\nu_{val})$in the $\nu$1f$_{7/2}$
and in $\nu$2p$_{3/2}$ are indicated at the top.}

\label{fig:fig1}
\end{figure}

In fig.\ref{fig:fig1} are shown, for N between 20 and 42, the known
experimental \bu values of the Ca, Ti, Cr, Ni and Zn isotopes. For N between 38
and 40, the \bu value decreases in the Ni but increases in the Zn isotopes.
This is explained in ref.~\cite{Sor01} as due to an increase of the deformation
with the neutron number N in the Zn isotopes. Consequently, the number of
single-particle levels in the neutron valence space increases which increases
the \bu. Moreover, the onset of deformation allows the quadrupole excitations
in $^{70,72}$Zn to avoid the hindering of their 1p-1h component due to the
parity of the $\nu 1g9/2$ orbital \cite{Sor01}. On the contrary, in the
spherical $^{68}$Ni (N=40, Z=28) the 1p-1h excitations are hindered and a very
low \bu value is observed \cite{Sor01}, 

\vspace{-0.75cm}  
\vspace{-0.5cm} 

\begin{figure}[ht]
\hspace{1cm} 
\includegraphics[scale=0.5]{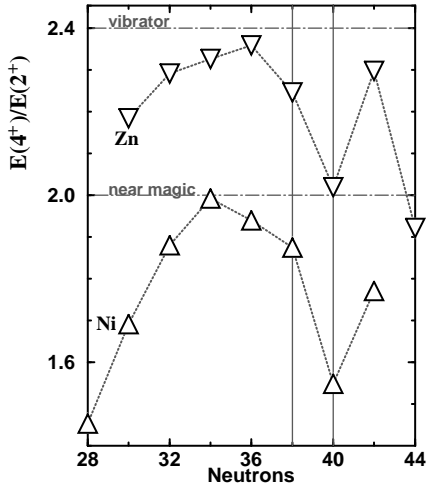} \vspace{-0.1cm}

\caption{ Experimental $E(4^+_1)/E(2^+_1$) values for the Ni and Zn isotopes
(dotted gray lines and symbols) obtained from the ENSDF file \cite{ENS};
near-magic and vibrators nuclei limits from \cite{Cas} (dot-dashed lines.)}

\label{fig:4ov2}
\end{figure}

In fig.\ref{fig:4ov2} we show the R$_4$=$E(4^+_1)/E(2^+_1$) ratios for the Ni
and the Zn isotopes. The near-magic and the vibrator limits of ref.\cite{Cas}
are also indicated. For N between 30 and 42 the striking parallelism of the Zn
and Ni curves indicates a constant difference of deformation between the Ni
isotopes and the Zn ones. For N between 38 and 40, the R$_4$ values are
decreasing, nearly down to the magic limit in the Ni isotopes and down to the
near-magic limit (R$_4$=2.0) in the Zn ones. Fig.\ref{fig:4ov2} indicates thus
a decreasing deformation in $^{64,68}$Ni and $^{66,70}$Zn. Even if one may have
reservations about this conclusion, because the R4 ratio brings into play two
more units of spin than the \buc, no doubt is possible about the rather small
deformation of $^{70}$Zn and $^{68}$Ni. For such nuclei the single-particle
levels evolve slowly as a function of the deformation (cf. a Nilsson diagram)
and there are as few single-particle levels going up and crossing the Fermi
level than going down. Therefore an increase of deformation in $^{66,70}$Zn --
unless it is a considerable one, which the Zn R$_4$ ratios totally exclude -- would
not increase the single-particle level density, no more than the \bu value. 
Another explanation has thus to be found.

\vspace{-0.6cm} 
\vspace{-0.6cm}

\noindent One can note in fig.\ref{fig:fig1} the common parabolic profile of
the \bu curves, as obtained in the seniority scheme, and for N between 32 and
38 the similarity of the Ni and Zn curves. This leads us to assume that protons
as well as neutrons are paired in the Ni as they are in the Zn isotopes and to
apply, in the search of new clues, the seniority scheme. It is worth noting
that to consider paired neutrons excludes that the neutrons 1p-1h excitations
could play in the \bu value the part assumed in \cite{Sor01}.\\ 

\vspace{-0.35cm}

In the seniority scheme, reducing the nucleus to N$_{val.}$ valence particles
(neutrons or protons) interacting in a single j-shell (1f$_{7/2}$ for example)
only by pairing, it can be shown \cite{R&S} that~:

\vspace{-0.3cm}
\begin{equation}
\text{\buc}_{seniority}\propto \text{\bua}
\end{equation}

\vspace{-0.6cm}
\begin{align}
\text{where\hspace{0.7cm}\bua}&=\frac{N_{val.}}{2}\times(\frac{\Omega_j}{2}-\frac{N_{val.}}{2})\\
&=N_{part.}\times N_{hol.}.
\end{align}
\vspace{-0.6cm}

\noindent $N_{part.}$ ($N_{part.}= \frac{N_{val.}}{2}$), $N_{hol.}$ and
$\Omega_j = (2j + 1)$ are the number of particle pairs, the number of hole
pairs and twice the number of levels in the single-j shell. The profile of \bua
is parabolic, with a maximum at mid-shell N$_{val.}$= $\Omega_j$/2 and two
symmetrical minima, one at the beginning of the shell N$_{val.}$=0, the other
at its end N$_{val.}$=$\Omega_j$. A known example of such a profile is given in
fig.\ref{fig:fig1} as a function of the $\nu_{val.}$ valence neutrons by the
\bu curve of the Ca isotopes series,which is a textbook case for the seniority
model \cite{Cas,Tal}. One can also see in fig.\ref{fig:fig1} that the Ni \bu
curve is, for N between 28 and 38, as regular as the Ca one. And yet, up to
three different neutron single j-shells are filled consecutively when going
from $^{58}$Ni to $^{68}$Ni. In first approximation we can then consider these
single-j shells as different members of one large and composite valence shell,
each contributing with the same weight to the \bu value. Having only in mind to
analyze the \bu curves in a very phenomenological way\footnote{It is important
to note that the extension of the single j-shell case to composite valence
shell, through the quasi-spin formalism (without weighting) or the generalized
seniority one, is the basis of the boson models \cite{IBM} widely used to
describe collective states.}, we will extend the expression of \bua obtained in
the seniority scheme to such a composite valence shell by replacing in (2)
$\Omega_j$ by \(\Omega = \sum \Omega_j\). This is equivalent to considering a
composite shell as a large single-j shell. In this approach the regular profile
up to N=38 of the Ni and Zn \bu curves (see fig.\ref{fig:fig1}) results from
the existence of a lower and an upper spacing between single-particle orbitals
(a gap if large or a sub-shell closure if small) confining the several single-j
shells in one valence shell. Any sharp discontinuity, such as in the Zn \bu
curve at N=38, can be interpreted as due either to the end of a neutron valence
at its upper spacing, or to the disappearance of one of the spacings. Indeed,
if the upper (or lower) spacing disappears, the valence shell will be enlarged
to the shell above (or below) it. Immediately $N_{hol.}$ (or $N_{part.}$) will
increase, the \bu value, related to their product, will suddenly be enhanced.
But such a change will also suddenly increase the deformation, and we have
excluded this. The Zn \bu curve highlights then a spacing at N=38. A careful
look to the experimental Ni \bu curve reveals one feature which can be also
related to a spacing at N=38~: the value at N=38 is nearly equal to the one at
N=28. It is worth noting that the \bu curves in the Ge (Z=32) and Se (Z=34), as
in the Zn isotopes, are increasing for N between 38 and 40 (and not decreasing
as in the Ni). For the Se, N=38 is even a minimum. The hypothesis of a N=38
sub-shell closure explains then more \bu features of this region than the one
of a doubly magic $^{68}$Ni. It remains to explain why the $^{68}$Ni \bu value
is the lowest one of the Ni series, N=28 and 38 included. To do so, within our
hypothesis, we need more information about the expected behavior of the \bu
curves of this region. In the next section we thus compare experimental \bu
curves with the ones calculated with various proton and neutron valence
spaces.

\vspace{-0.5cm}
\section{\label{sec:calc}Comparison with calculations}

We take into account both type of particles by summing the proton and the
neutron contributions. Replacing in eqs.(3) N$_{part.}$, N$_{hol.}$ by
$\nu_{part.}$, $\nu_{hol.}$  for the neutrons particle and hole pairs, and by
$\pi_{part.}$, $\pi_{hol.}$ their equivalents in proton, we have computed
the following expression~: 

\vspace{-0.5cm} 
\begin{equation} 
\text{\buc}_{gen.}= [(\nu_{part.}\times\nu_{hol.}) +(\pi_{part.}\times\pi_{hol.})]
\end{equation} 

\noindent  where $\nu_{part.}$, $\nu_{hol.}$, $\pi_{part.}$ and $\pi_{hol.}$
are calculated accordingly to the proton and neutron valence spaces. 

\begin{figure}[h]   
\hspace{-0.3cm}    
\includegraphics[scale=0.5]{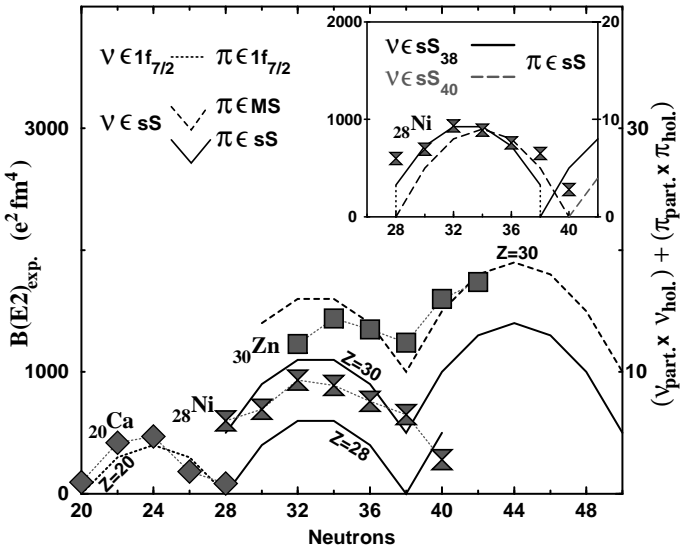}

\caption[]{Experimental Ca, Ni and Zn \bu curves from \cite{Ram2} (symbols, 
dotted gray lines) and calculated ones with different proton and neutron
valence spaces (see text and legend). Inset: experimental Ni \bu values and
calculated profiles with a proton sub-shell closure and a neutron one located
either at N=38 (shifted curve, black line) or at N=40 (dashed gray line).}

\label{fig:calculs} 
\end{figure}

\noindent We have performed the calculations for Z=28 and 30 (varying N between
28 and 50) and also for Z=20 (20$\leq$N$\leq$28). These last ones are used to
normalize our results to the $_{20}$Ca experimental values. In the Ca isotopes
there is only one possibility for the proton and neutron valence space, the
$\pi$1f$_{7/2}$ and $\nu$1f$_{7/2}$ single-j shells. On the contrary, when N,
as Z, are above 28 (neutrons in $^{56-68}$Ni and $^{62-72}$Zn and protons only
for the $_{30}$Zn isotopes) it exists two possibilities of valence space~:
either the full 28-50 major shell (MS), or a sub-shell (sS) resulting from a
sub-shell closure. As previously mentioned, the sub-shell closure is assumed to
take place at N=38 for neutrons. For protons, accordingly to the \bu ratio of two
consecutive isotones and to the \bu curves of isotonic series, it is assumed at
Z=40 \cite{Del03}. The neutron and proton valence spaces used to calculate the
curves, drawn with different lines in fig.\ref{fig:calculs}, are given in the
legends. The Z=28 curve and the two Z=30 ones present a minimum at N=38 (as
stressed by the legend) accordingly to the N=38 sub-shell closure assumption. A
major shell for the neutron valence space would have given a maximum for
N=38-40 (see ref.\cite{Del03}).     

\vspace{-0.5cm}     
\vspace{-0.5cm}

\noindent These calculations provide a frame of ``smooth behaviors"
highlighting the significant deviations of the experimental curves.
Fig.\ref{fig:calculs} makes evident that the experimental Ni curve lies higher
than expected, very near the lowest Z=30 curve. As proton magic nuclei, their
\bu should be null for N=28, like for the $^{48}$Ca, and the other values
should lie on the Z=28 curve which prolongates the Z=20 one. The Ni
experimental \bu value at N=40 which was considered ``abnormal" since lower
those at N=28 and N=38, is in fact the only normal one, near the Z=28
calculated value. All other Ni \bu values are too high, increased from N=28 to
N=38 by a phenomenon disappearing at N=40. This is more probably due to the
protons than to the neutrons. Indeed, a neutron effect would evolve with N and
change the shape of the Ni \bu curve. But, in the insert of
fig.\ref{fig:calculs} one can see the perfect agreement along four points for N
between 30 and 36 of the experimental profile and the one calculated with a
N=38 neutron sub-shell closure (sS$_{38}$, solid black line). This latter curve
has been shifted up to underline the agreement and to illustrate the hypothesis
of a constant enhancement. The curve obtained under the assumption of a N=40
neutron sub-shell closure (sS$_{40}$, dashed gray line) exhibits indeed another
profile and does not allow to obtain the previous agreement. The enhancement
appears mainly due to the protons, except at N=28 and N=38 were the slight
disagreement can be related to a neutron effect, as discussed in \cite{Del03}.
On the other hand, in the insert of fig.\ref{fig:calculs}, one can note that
the N=40 neutron sub-shell closure assumption gives a null \bu value for
$^{68}_{28}$Ni, lower than the experimental one. Therefore, the hindering of
the quadrupole excitation assumed in \cite{Sor01} would go in the wrong way for
reproducing the $^{68}$Ni data in a N=40 sub-shell closure hypothesis. The
decrease in the Ni isotopes between N=38 and 40 is really misleading, it
dissimulates a N=38 neutron sub-shell closure behavior and a proton effect. 

\noindent The agreement between experiment and calculations is more visible in the Zn
(excellent above N=34) than in the Ni isotopes thanks to the experimental
minimum at N=38. Nevertheless, between N=32 and 34 the Zn experimental curve
passes from near the curve obtained with proton and neutron sub-shell closures
to the curve obtained with a major-shell for the proton valence space, where it
stays.

\indent In both Ni and Zn \bu curves, the proton contribution is increased for
some neutron numbers. The change of the proton valence space between
the lightest Zn isotopes and the heaviest ones can be interpreted as the
vanishing of the proton sub-shell closure. All this calls to mind a
proton-neutron interaction effect. Indeed, it has been pointed out
\cite{F&P,Cas81} that the p-n interaction can be strong enough to reduce (and
even to eradicate) the Z=40 sub-shell gaps in the heavy (N$\geq$60) Zr, Mo and
Ru isotopes \cite{F&P,Cas81,Ciz97}, this gap reduction going with promotion of
protons over it and of neutrons into higher orbitals \cite{F&P,Cas81,Ciz97}.  

\vspace{-0.5cm}
\section{\label{sec:Ni}Anchors of our interpretation} 
\vspace{-0.2cm}

\begin{figure} 
\includegraphics[scale=0.45]{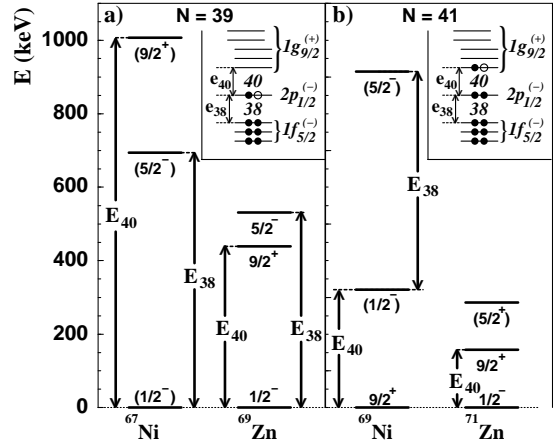}
\vspace{-0.2cm}
\caption{Experimental excitation energies from \cite{ENS} of the first three
states in a) $^{67}$Ni and $^{69}$Zn, b) $^{69}$Ni and $^{71}$Zn. Insets:
Neutron single-particle level order from \cite{Naz85} and occupation for the
ground states.}
\vspace{-0.6cm}
\label{fig:gaps} 
\end{figure}

Our interpretation of the Ni and Zn \bu curves is built firstly on a neutron
sub-shell closure at N=38. A N=38 spacing has been obtained between the $1f_{5/2}$
and $2p_{1/2}$ orbitals in \cite{Naz85}, as shown in the insets of 
fig.\ref{fig:gaps}. One can note that this single-particle level spectrum
offers at N=40 the possibility for a second energy spacing. 

\noindent We involve also in our interpretation the p-n interaction. It is
worth noting that the 1f$_{7/2}$ (for particle number ranging between 22 and
28), 1f$_{5/2}$ and 1g$_{9/2}$ orbitals surrounding N=38 and 40 can allow the
p-n interaction to start \cite{Del03}, since they make possible a large overlap
between the proton and neutron orbitals \cite{DSh53}. The N=40 spacing, if
existing for the Ni isotopes only, could complete our interpretation of the 
difference between the \bu curves in $^{66,68}$Ni and $^{68,72}$Zn. In addition
to the Z=28 gap, two consecutive spacings between $\nu$1f$_{5/2}$ to
$\nu$1g$_{9/2}$ would represent an insuperable obstacle for the p-n interaction
unable to act in the Ni isotopes as it does in the Zn isotopes after N=38.

\indent The presence of a N=38 neutron sub-shell closure in both the Ni and Zn
isotopes followed by, in the Ni isotopes only, a N=40 energy spacing is
confirmed by the energies of the two first excited states in the Ni and Zn odd
isotopes with N=39 (drawn in fig.\ref{fig:gaps}a) and 41 (fig.\ref{fig:gaps}b).
In fig.\ref{fig:gaps} we have called E$_{38}$ and E$_{40}$ the difference of
excitation energy related respectively to e$_{38}$, the N=38 spacing between
$1f_{5/2}$ and $2p_{1/2}$, and to e$_{40}$, the N=40 one between $2p_{1/2}$ and
$1g_{9/2}$. We can see on fig.4 that e$_{38}$ and e$_{40}$ are both active in
the odd-Ni isotopes : E$_{38}$ and E$_{40}$ are so large that the $^{67,69}$Ni
excitation spectra, with only 2 excited states below 1MeV (900keV for ${69}$Ni)
have twice smaller density than in their corresponding odd-Zn isotopes. Indeed,
in $^{69}$Zn E$_{38}$ is 100keV smaller and E$_{40}$ is twice smaller than in
$^{67}$Ni\footnote{The decrease of E$_{38}$ and E$_{40}$ from Ni to Zn
odd-isotopes is also probably due to the p-n interaction}. In $^{71}$Zn the
$1g_{9/2}$ orbital is so close to $2p_{1/2}$ that 1f5/2 (too far) is not
anymore involved in the first excited states; E$_{38}$ can not anymore be
deduced. From the N=40 spacing, the \bu value of the even $^{70}$Ni can be
expected around 400e$^2$fm$^4$ (on the sS$_{40}$ curve of the inset of
fig.~\ref{fig:calculs}), near the $^{68}$Ni \bu value. On another hand, in
$^{70}$Ni two neutrons are occupying $\nu1g_{9/2}$. The p-n interaction could
restart, eradicate the N=40 sub-shell closure. The N=38 sub-shell closure would
then be the lower limit of the neutron valence space and the $^{70}$Ni \bu
value will lie around 800e$^2$fm$^4$ (on the sS$_{38}$ curve of the inset) or
even above if the proton promotion restarts also. It is worth noting that the
$^{68}$Ni \bu value lies exactly between the two curves of the inset as a
transitional point would. Recent lifetime measurements \cite{kenn} give lower
$^{58-64}$Ni \bu values than in \cite{Ram2}. If confirmed, such values --which
are in agreement with a N=38 sub-shell closure without even involving proton
contribution -- put with the values at N=28 and 38, which are in agreement with
N=40 sub-shell closure, would extend this transitional behavior to the whole Ni
isotopic series. 

\indent The Ni and Zn \bu curves are analyzed using calculations performed
within the seniority scheme. They are shown to follow a N=38 sub-shell closure
behavior with deviations interpreted as coming from a proton contribution
enlarged by the p-n interaction. The low \bu value in $^{68}$Ni is explained by
the Z=28 gap which, added to the N=38-40 sub-shell spacings, prevents the p-n
interaction to promote protons over the Z=28 gap (as it does in $^{56,66}$Ni)
together with (as in $^{64,72}$Zn) neutrons over the N=40 energy spacing. The
$^{70}$Ni \bu value is expected either near the $^{68}$Ni one or twice larger
if the p-n interaction can restart despite the Z=28 and N=38-40 gaps. Indeed,
in $^{66,72}$Zn the p-n interaction is found strong enough to suppress the Z=40
proton sub-shell closure. The \bu value in the heavier even Zn isotopes is
expected to increase up to N=44 then to decrease.

\vspace{-0.1cm}
\vspace{0.1cm}
\vspace{-0.3cm}
\acknowledgement{The authors want to thank R. Lombard for discussions and J.
Kiener, J. Sauvage, and D. Lunney for attentive readings.} 


\vspace{-0.4cm}
\begin{thebibliography}{10}
\bibitem{Sor02}
O.~Sorlin \it{et~al.}\n , Phys.
Rev. Lett. \textbf{88} 092501 (2002).

\bibitem{Sor01}
O.~Sorlin \it{et~al.}\n ,
  E. P. J. \textbf{14} 1 (2002).

\bibitem{Sor02b}
O.~Sorlin, Private Communication (2002).

\bibitem{May02}
P.~Mayet,\it{www.cern.ch/ISOLDE/Workshop2003/mayet.ppt}\n

\bibitem{Lan03}
K.~Langanke, J.~Terasaki, F.~Nowacki, D.J. Dean and W.~Nazarewicz, Phys.
  Rev. C \textbf{67} 044314 (2003).

\bibitem{Cau04}
E.~Caurier, G.~Martínez-Pinedo, F.~Nowacki, A.~Poves, A.~P.~Zuker arXiv:nucl-th/0402046 ~1 (2004).

\bibitem{Del03}
I.~Deloncle and B.~Roussi\`ere arXiv:nucl-th/0309050 ~1 (2003),
  and to be published.

\bibitem{Ram2}
S.~Raman, C.W.~Nestor, J.R. and P.~Tikkanen, At. Data and Nucl. Data Tab.
  \textbf{78} 1 (2001).

\bibitem{Cas}
R.F. Casten \emph{Nuclear Structure from a Simple
  Perspective}\n (Oxford University Press,
  1990).

\bibitem{ENS}
\emph{Evaluated Nuclear Structure Data File}\n  maintained by the National
  Nuclear Data Center, Brookhaven National Laboratory.

\bibitem{R&S}
P.~Ring and P.~Schuck \emph{The nuclear many body problem}\n 
  (Springer-Verlag New York Inc, 1980).

\bibitem{Tal}
I.~Talmi \emph{Simple Models of Complex Nuclei} vol.~7 of \emph{Contemporary
Concepts in Physics} (Harwood Academic Publishers, 1993).

\bibitem{IBM}
F.~Iachello ed., \emph{Interacting Bosons in Nuclear Physics}\n  vol.~1 (Plenum
  Press, New York and London, 1978).

\bibitem{F&P}
P.~Federman and S.~Pittel Phys. Lett. \textbf{69B} 385 (1977).

\bibitem{Cas81}
R.F.~Casten, D.D.~Warner, D.S.~Brenner and R.L.~Gill, Phys. Rev. Lett. \textbf{47} 1433
  (1981).

\bibitem{Ciz97}
W.~Younes  and J.A.~Cizewski, Phys. Rev. C \textbf{55} 1218 (1997).

\bibitem{Naz85}
W.~Nazarewicz, J.~Dudek, R.~Bengstson, T.~Bengstson, I.~Ragnarsson, Nucl. Phys.
  A \textbf{435} 397 (1985).

\bibitem{DSh53}
A.~De-Shalit and M.~Goldhaber Phys. Rev. \textbf{92} 458
  (1953).
\bibitem{kenn}
O.~Kenn \it{et al.}\n Phys. Rev. C \textbf{63} (2000) 21302

\end{thebibliography}
\end{document}